\documentclass[preprint,,amsmath,amssymb]{revtex4-2}

\pdfoutput=1
\usepackage{graphicx}
\usepackage{color}
\usepackage{dcolumn}
\usepackage{bm}

\begin{document}


\title{Spherical symmetric dust collapse in a Vector-Tensor gravity}

\author{Roberto Dale}
\email{rdale@umh.es}  
\affiliation{
Departamento de Estad\'{\i}sica, Matem\'atica e Inform\'atica, 
Universidad Miguel Hernandez, Elche, Alicante, Spain    \\
\\
Center of Operations Research (CIO),
University Miguel Hernandez of Elche (UMH)\\
}%

\author{Diego S\'aez}
\email{diego.saez@uv.es}
\affiliation{Departamento de Astronom\'{\i}a y Astrof\'{\i}sica, Universidad de Valencia,
46100 Burjassot, Valencia, Spain.  \\ 
\\
Observatorio Astron\'omico, Universidad de Valencia, 
E-46980 Paterna, Valencia, Spain\\
}%

\date{\today}

\begin{abstract}
There is a viable vector-tensor gravity (VTG) theory,
whose vector field produces repulsive forces leading to 
important effects. In the background universe, 
the effect of these forces is an accelerated expansion identical 
to that produced by vacuum energy (cosmological constant). 
Here, we prove that another of these effects 
arises for great enough collapsing masses which lead to Schwarzschild
black holes and singularities in general relativity (GR). For these masses, 
pressure becomes negligible against gravitational attraction and 
the complete collapse cannot be stopped in the context of GR; however, in VTG,
a strong gravitational repulsion could stop the falling of 
the shells towards the symmetry center. A certain study of a collapsing dust cloud
is then developed and, in order to undertake this task, the VTG equations in comoving coordinates are written.
In this sense and, as it happens in general relativity for a pressureless dust ball, three different solutions are found.
These three situations are analyzed and the problem of the shell crossings is approached. 
The apparent horizons and trapped surfaces, whose analysis will lead to diverse situations, depending on certain theory characteristic parameter value, are also examined.

\end{abstract}



\maketitle

\section{Introduction}
\label{sec:1}

Any vector-tensor theory of gravitation involves the metric tensor $g^{\mu \nu} $
and a vector field $A^{\mu} $. These fields are coupled to build up an appropriate
action leading to the basic equations via variational calculations.
There are many actions and vector-tensor theories \citep{Wil93,Wil06}, 
but one of them has been extensively 
studied to conclude that: (i) it has not either classical or quantum instabilities and,
(ii) it explains -as well as general relativity (GR)- both
cosmological and solar system observations 
\citep{Dal09,Dal12,Dal14,Dal15,Dal17}; hence,
new applications of this viable and promising theory are worthwhile. 
Since there are opposite gravitational forces, this theory
will be hereafter called AR-VTG (attractive-repulsive vector-tensor gravity).                     

As it was shown in \cite{Dal15}, for appropriate values of the AR-VTG parameters (see below),
there are black hole event horizons with admissible radii which are a little smaller than 
those of GR; nevertheless, for other values of these parameters, there are no horizons of 
this kind. 

Our signature is (--,+,+,+). Greek 
indexes run from 0 to 3. Symbol $\nabla $ ($\partial $) stands
for a covariant (partial) derivative. The antisymmetric tensor $F_{\mu \nu} $
is defined by the relation 
$F_{\mu \nu} = \partial_{\mu} A_{\nu } - \partial_{\nu} A_{\mu }$. It has nothing to 
do with the electromagnetic field. 
Quantities $R_{\mu \nu}$, $R$, and $g$ are the covariant components of the Ricci 
tensor, the scalar curvature, and
the determinant  of the matrix $g_{\mu \nu}$ formed by the covariant components 
of the metric, respectively. 
Units are chosen in such a way that the gravitational constant, $G$, and the speed of light, $c$,
take the values $c= G = 1$; namely, we use 
geometrized units.
 
This paper is structured as follows: The AR-VTG theory is described in Sect. \ref{sec:2}; 
the vacuum stationary spherically symmetric solutions of the field equations are 
presented in Sect. \ref{sec:3}; the collapsing systems to be considered are 
described in Sect. \ref{sec:4}; the behavior of the collapse of a spatially bounded spherical dust cloud, 
modelled into shells, is considered in Sect. \ref{sec:7}, and finally, 
Sect. \ref{sec:8} contains conclusions, a certain discussion, and prospects. 

\section{AR-VTG foundations}
\label{sec:2}

Let us now briefly summarize
the AR-VTG basic equations, which were 
derived in \cite{Dal09,Dal12} from an appropriated 
action, which is a particularization of the general vector-tensor action 
given in \cite{Wil93}. The resulting field equations are:
\begin{equation}  
G^{\mu \nu} = 8\pi (T^{\mu \nu}_{GR} + T^{\mu \nu}_{VT}) \ ,
\label{fieles_vt}
\end{equation}   
\begin{equation}
2(2\varepsilon - \gamma)\nabla^{\nu} F_{\mu \nu} = J^{^{A}}_{\mu} \ ,
\label{1.3_vt}
\end{equation}    
where $G^{\mu \nu}$ is the 
Einstein tensor, $T^{\mu \nu}_{GR}$ is the GR energy momentum tensor,
$J^{^{A}}_{\mu} \equiv -2 \gamma \nabla_{\mu} (\nabla \cdot A)$ 
with $\nabla \cdot A = \nabla_{\mu} A^{\mu} $, and 
\begin{eqnarray}
T^{\mu \nu}_{VT} &=& 2(2\varepsilon - \gamma) [F^{\mu}_{\,\,\,\, \alpha}F^{\nu \alpha}
- \frac {1}{4} g^{\mu \nu} F_{\alpha \beta} F^{\alpha \beta}] \nonumber \\ 
& &
-2\gamma [ \{A^{\alpha}\nabla_{\alpha} (\nabla \cdot A) + \frac {1}{2}(\nabla \cdot A)^{2}\}
g^{\mu \nu} \nonumber \\
& &
-A^{\mu}\nabla^{\nu} (\nabla \cdot A) - A^{\nu}\nabla^{\mu} (\nabla \cdot A)
] \ .
\label{emtee_vt}
\end{eqnarray}  
Equation (\ref{1.3_vt}) leads to the following conservation law 
\begin{equation}
\nabla^{\mu} J^{^{A}}_{\mu} = 0  
\label{confic}
\end{equation}
for the fictitious current $J^{^{A}}_{\mu}$. Moreover, the 
conservation laws $\nabla_{\mu} T^{\mu \nu}_{GR} = 0 $ and  
$\nabla_{\mu} T^{\mu \nu}_{VT} = 0 $ are satisfied by any solution 
of Eqs.~(\ref{fieles_vt}) and~(\ref{1.3_vt}) (see \cite{Wil93}).

The pair of parameters
($\varepsilon$, $\gamma$) must satisfy the inequality $2\varepsilon - \gamma > 0$ to prevent 
the existence of quantum ghosts and unstable modes in AR-VTG (see \cite{Dal14} an references 
cited therein). 
The condition $\gamma >0$ must be required to have a positive $A^{\mu} $ energy density 
in the background universe, which will play the role of vacuum energy; hence, 
the inequalities 
$\varepsilon > \frac {\gamma}{2}>0 $ must be satisfied.

\section{Vacuum stationary spherically $\,\,\,$ symmetric metrics in AR-VTG}   
\label{sec:3}

In the stationary spherically symmetric case, in Schwarzschild coordinates, the line element
may be written as follows [see e.g., \cite{Ste03}]:
\begin{eqnarray}
ds^{2}&=&\ - e^{\nu} dt^2 + e^{\lambda} dr^2 \nonumber \\
& &
+ r^2 ( \ d\theta^2 + sin^2\theta \ d\phi^2) \ , 
\label{LE}
\end{eqnarray}   
where $\nu$ and $\lambda $ are functions of $r$. Moreover, the covariant components $A_{\mu}$ have the form:
\begin{equation}
A_{\mu} \ \equiv \ (A_0,A_1,0,0) \ .
\label{VF}
\end{equation}
In the absence of any 
matter content, the form of
functions $\nu(r)$, $\lambda(r)$, $A_0(r)$, $A_1(r)$, and $\nabla \cdot A$
may be found in \cite{Dal15}. The resulting functions involve integration constants and
some of them have not yet been fixed; nevertheless, the involved constants  in 
$A_0(r)$, and $A_1(r)$ are not necessary to perform
this research since, whatever their values may be, the AR-VTG line element is:
\begin{eqnarray}
ds^{2}&=&\ - h(r) dt^2 + h^{-1}(r) dr^2 \nonumber \\
& &
+ r^2 (d\theta^2 + sin^2\theta \ d\phi^2) \ , 
\label{LEK}
\end{eqnarray}      
where 
\begin{equation}
h(r) = 1-\frac{2M}{r}-\frac {\Lambda}{3}r^{2} + \frac {\alpha^{2}M^{2}}{r^{2}} \ .
\label{rnm}
\end{equation}   
The dimensionless quantity $\alpha^{2} > 0$ is proportional to 
the positive number $ 2 \varepsilon - \gamma $. 
It is obvious that the metric (\ref{rnm}) is formally identical to the Reissner-Nordstr\"{o}m-de 
Sitter metric of Einstein-Maxwell theory for a stationary spherically symmetric charged system,
with charge $Q$
such that $Q^{2} = \alpha^{2} M^{2} $. For $Q=0$ this metric reduces to 
the
Kottler-Schwarzschild-de Sitter metric \citep{Kot18} of GR. 

Finally, as it was shown in \cite{Dal15}, the scalar $\nabla \cdot A$ is constant.
Its value may be easily fixed taking into account Eq.~(\ref{emtee_vt}). In fact, 
according to this equation, as $r$ tends to infinity, the energy momentum tensor $T^{\mu \nu}_{VT}$
created by the total mass $M$ tends to $-\gamma (\nabla \cdot A)^{2} \eta^{\mu \nu}$, 
where $\eta^{\mu \nu}$ is the Minkowski metric.
Since this energy momentum tensor must asymptotically vanish, one concludes that the relation
$\nabla \cdot A =0$ must be satisfied for the vacuum solution under consideration.
This relation and the fact that $M$ is a constant will be taken into 
account to fit the inner and outer solutions on the collapsing star boundary.

\section{General considerations}   
\label{sec:4}

In terms of the function $h(r) = - g_{00}(r)$,
the event horizons are the hypersurfaces $r=r_{h} $ defined by the condition $h(r_{h}) =  0$.
A complete discussion about horizons -for the line element defined by Eqs.~(\ref{LEK}) and~(\ref{rnm})--     
may be found in \cite{Kay79}. The number of 
horizons and their radius depend on the values of $M$, $\Lambda$ and $\alpha^{2} $.

In the standard $\Lambda$CDM cosmological model of GR, most current observations are explained 
for a vacuum energy density parameter $\Omega_{\Lambda} \simeq 0.7$; namely, for 
$\Lambda \simeq 1.2 \times 10^{-46} \ km^{-2} $. The same value 
also explains current observations in the framework of AR-VTG (see \cite{Dal14}); hence,
this value of the cosmological constant is hereafter fixed.
In AR-VTG cosmology, quantity $\nabla \cdot A $ takes on a constant value, and 
the cosmological constant is proportional to the square of this value; nevertheless,
this constant must have another 
unknown origin in GR;
anyway, in the stationary spherically symmetric case, it is 
denoted $\Lambda$ and its treatment is the same in both theories.
The line element (\ref{LE}) does not depend on time and, consequently, 
it does not describe a cosmological space-time, but the space-time 
in a region located well inside the so-called cosmological horizon
and outside the black hole event horizon (if it exists). In this region one has 
$h(r)>0 $ as it should be. Sometimes, the black hole horizon does not exist 
and the metric is well defined everywhere -inside the 
cosmological horizon- excepting the singular point $r=0$.

For realistic masses, $M$, ranging from those of the smallest black holes 
(star collapses) to the masses of the greatest 
super massive black holes located in galactic centers, the product
$\Lambda M^{2} $ is many orders of magnitude smaller than unity and,
as a result of this fact, the general discussion about horizons presented 
in \cite{Kay79} may be simplified. First of all, it may be easily proved that,
for arbitrary realistic values of $M$ and $\alpha^{2}$ (see below), there is a 
cosmological horizon with radius $r_{c} \simeq 1.73 \times 10^{23} \ km $;
this is the outermost horizon, whose radius essentially depends on the cosmological 
constant, which
has been fixed; furthermore, for $r$ of the order of $M$ ($r << r_{c}$), 
other inner horizons may
exist. The radius of these horizons may be easily estimated, with very high precision,
by neglecting the term $\Lambda r^{2} /3$ -order $\Lambda M^{2} $- in the $h(r) $ formula; 
the error due to this approximation is fully negligible. By solving then the equation 
$h(r) =  0$ one easily finds: (i) for $\alpha^{2} > 1$, there are no solutions and,
consequently, only the cosmological horizon exists, (ii) for $\alpha^{2} = 1$,
there is an unique double solution $r=M$ corresponding to 
an inner event horizon and, (iii) for 
$\alpha^{2} < 1$ there are two solutions $r_{H_{-}}$ and $r_{H_{+}}$ with $r_{H_{-}} < r_{H_{+}}$.
In such a situation, there are two inner horizons plus the cosmological one.

In the context of GR, after a supernova explosion, three cases may be distinguished: 

(1) if the mass $M$ of the supernova core is smaller than the Chandrasechar limit ($\sim 1.4 M_{\odot}$), 
a white dwarf is formed. As an example, let us mention Sirius B \citep{Hol98}, with a mass $M \sim M_{\odot}$ and a radius
$R \sim 8\times 10^{-3} R_{\odot}$, this white dwarf has a ratio $M/r \sim 3\times 10^{-4}$ 
and a mean density $\rho \sim 3\times 10^{6} \ gr/cm^{3} $; in this situation,
the pressure of degenerate electrons prevents collapse, and the value of $M/r$ is so small 
that the term $\alpha^{2}M^{2}/r^{2}$ in Eq.~(\ref{rnm}) is negligible against $2M/r$
for $\alpha^{2} \leq 2$, 
which means that AR-VTG, with $\alpha^{2} \leq 2$ (see below), and 
GR ($\alpha^{2}=0$) lead to
the same description of Sirius B. The same occurs for any white dwarf, 

(2) for a mass $M \gtrsim 1.4 M_{\odot}$, the pressure due to degenerate electrons cannot 
balance gravity and star contraction continues to reach densities much greater than 
$\rho \sim 10^{6} \ gr/cm^{3} $. For masses $1.4 \lesssim M \lesssim 3 \ M_{\odot}$ 
\citep{Sha04}, the pressure due to degenerate neutrons and their strong interactions
may prevent collapse to form a dense neutron star with a small radius 
of the order of $10 \ km$ \cite{Oze16}; e.g., a great neutron star, 
with mass $M = 2 M_{\odot}$ and radius $R \sim 10 \ km$, 
has a ratio $M/r \sim 0.15$ and a mean density $\rho \sim 2 \times 10^{15} \ gr/cm^{3} $ 
(see \cite{Mis73}). For this value of $M/r$ and $\alpha^{2} \leq 2$, the term 
$\alpha^{2}M^{2}/r^{2}$ is less than 15\% of $2M/r$; hence, small but non-negligible
deviations between the neutron star structures in GR and AR-VTG should exist.
Since these deviations could be too great for $\alpha^{2} >> 2$, in this paper, 
it has been tentatively assumed the condition $\alpha^{2} \leq 2$.
The mass-radius relation must be estimated in the context of AR-VTG; namely, taking into account the
$A^{\mu} $ repulsion for different $\alpha^{2} $ values. This will be done elsewhere,

(3) for masses $M \gtrsim 3 M_{\odot}$, the pressure of degenerate neutrons 
cannot prevent collapse. All the particles would reach the singularity $r=0$ 
in a finite proper time, although this fall towards $r=0$ cannot be observed 
from points with $r > 2M $ 
due to the existence of an event horizon at $r=2M$. As the system approaches the singularity,
its density grows, reaching greater values  than the so-called Planck density. 
For these huge densities, classical gravity does not apply and quantum gravity might prevent 
the singularity; however, as it will show in next sections, the shells of a collapsing star 
cannot reach the singularity ($r=0$) in AR-VTG due to the action of a 
strong $A^{\mu} $ repulsion. Nothing similar occurs in GR (just gravitation, that is, no external fields)
where any neutral  (not charged) shell collapses.

The question is: what happens with star and black hole formation in the context of AR-VTG 
for $\alpha^{2} \leq 2$? A qualitative but accurate answer follows from points 
(1)-(3). For $M \lesssim 3 \ M_{\odot}$, GR and AR-VTG predict similar 
qualitative evolutions, even for the most massive observed 
neutron stars. The differences between both theories
grow as $M$ increases.
For $M \gtrsim 3 \ M_{\odot}$, 
the nuclear density $\rho_{N} \sim 2 \times 10^{14} \ gr/cm^{3} $ will be  
reached at a certain moment, $t_{N}$, as in neutron stars, but the pressure of degenerated nucleons
cannot stop contraction when this density is reached and, consequently, collapse continues. 
Let us study this last phase starting at time $t_{N}$. 
Close to this time RG and AR-VTG evolutions would be comparable, but 
these evolutions dramatically deviate later. 
In GR, contraction cannot be stopped and particles converge towards
$r=0$; however, in the context of AR-VTG, it may be proved that
the repulsive gravitational forces quickly grow as $r$ decreases, in such a way that,
the fall of the fluid shells towards $r=0$  
is stopped when certain minimum bounce radius
is reached. Nevertheless, specifically because of this increasing repulsive force, 
the innermost shells are gravitationally less bounded, and hence, collapse more slowly than the outer ones, 
and, as it happens with an electromagnetic repulsive force for a collapsing charged dust, shell crossings is a possibility 
that have to be reviewed; this is complemented with an analysis about the formation of horizons and trapped surfaces.
The determination of the conditions under which trapped surfaces are formed (if there exist) is an interesting, or
furthermore necessary, task because it may reveal the presence of singularities in a gravitational collapse \cite{Pen65}.
Analytical models for the evolution of the dust collapsing cloud is considered in next section,
where the aforementioned issues are also examined.

\section{Singularities in GR and AR-VTG}   
\label{sec:7}

In the framework of GR, any shell -- including the outermost one -- (see \cite{Lan75})
of the pressureless collapsing dust, reaches point $r=0$ in a proper finite 
time $\Delta \tau $, thus, the collapse of the whole dust cloud is
unavoidable. However, in AR-VTG, the outermost shell has a very different 
behavior. One could expect that the dust sphere's surface moves as a test particle in the AR-VTG stationary
spherically symmetric spacetime defined by the line element given by Eqs.~(\ref{LEK}) and~(\ref{rnm}).
Neglecting the term $\Lambda r^2/3$, the radial motion is ruled by the equation:

\begin{equation}  
\frac {d\tau}{dx} =  \pm M \Big[ (E^2 - 1)  + \frac {2}{x} -
\frac {\alpha^2} {x^2} \Big]^{-1/2}  \ ,
\label{proptev}
\end{equation}
where $E$ is the energy per unit rest mass, $\tau$ is the proper time and $x$ is defined
by the ratio $x \equiv r/M$. The causal structure of this spacetime is analogous to the 
Reissner-Nordstr\"{o}m one (see figure 1 in \cite{Ori91}) for $\alpha^2 < 1$, and
the apparent horizons are defined by the expression:

\begin{equation}  
r_{H_\pm} = M ( 1 \pm \sqrt{ 1 - \alpha^2} )  \ .
\label{ARVTGHor}
\end{equation}
  
A possible solution of Eq.~(\ref{proptev}) would lead to damped oscillations with a small period driving the test particle
to a state of minimum gravitational energy with a finite radius. Point 
$r=0 $ would be not reached by the aforementioned test particle; 
this situation is due to the existence of the repulsive component of the AR-VTG
modified gravity. As it happens in a collapsing charged dust (see \cite{Ori91}),   
the motion of the outermost shell may not be independent of the evolution of the
inner dust cloud. For instance, in the same way that in \cite{Ori91}, the shell crossings may break down prior expectations.
This possibility has to be analyzed.

Hereafter, it is assumed that the collapsing core 
is big enough and pressure gradients become negligible against gravitational 
forces. This core is an ideal fluid with $T^{\mu \nu}_{GR} = \rho u^{\mu} u^{\nu}$,
where $\rho$ is the energy density, $u^{\mu} = dx^{\mu}/d\tau$ the 4-velocity,
and $\tau $ the proper time. The non vanishing components of the 4-velocity 
are $u^{0} $ and $u^{1} $.

\subsection{The Basic Equations in Schwarzschild-like coordinates}
\label{sec:7a}

Let us first consider Schwarzschild-like coordinates to write the interior non stationary
spherically symmetric metric in the form (\ref{LE}), with $\nu=\nu(r,t)$ and 
$\lambda=\lambda(r,t)$. In this way, the interior solution may be matched with the 
exterior vacuum solution given by Eqs.~(\ref{LEK}) and~(\ref{rnm}).

It has been proved (see above) that the scalar $\nabla \cdot A$ vanishes outside the 
collapsing core. Let us now assume that, in the spherically symmetric case, this scalar
vanishes everywhere (also inside the collapsing object); 
so, there are no matching problems with $\nabla \cdot A$ at the core surface. This assumption 
will be proved to be 
consistent with the AR-VTG equations and it is necessary to coherently match 
the inner and outer solutions (see below).
In spite of the fact that $\nabla \cdot A $ vanishes everywhere, functions $A_0 $ and $A_1 $
do not simultaneously vanish and AR-VTG does not coincide with GR in the 
spherically symmetric case (spherical collapse).

For $\nabla \cdot A = 0$, the field equations (\ref{1.3_vt}) reduces to
\begin{equation}
\nabla^{\nu} F_{\mu \nu} = \frac {1}{\sqrt{-g}} \partial (\sqrt{-g} F^{\mu \nu})/\partial x^{\nu} = 0 \ . 
\label{1.3_vtm}
\end{equation}    

Inside the collapsing object, the non vanishing components of the AR-VTG vector field 
are $A_0 = A_0(r,t)$ and $A_1 = A_1(r,t)$ and, the nonvanishing components of $F_{\mu \nu}$
are $F_{01} = -F_{10} = \partial A_{1} /\partial t - \partial A_{0} /\partial r$. It is then easily 
proved that Eqs.~(\ref{1.3_vtm}) reduce to
\begin{equation}
\partial (r^{2} e^{\sigma} F^{01})/ \partial r= 0  
\label{1.3_vtm_a}
\end{equation}    
and
\begin{equation}
\partial (r^{2} e^{\sigma} F^{01})/ \partial t= 0 \ , 
\label{1.3_vtm_b}
\end{equation}    
where $\sigma = (\lambda + \nu)/2$; hence,
$r^{2} e^{\sigma} F^{01}$ quantity is a constant. 

In the absence of electrical charges and currents,
Eq~(\ref{1.3_vtm}) is also valid in Einstein-Maxwell theory.
In such a case, the relation 
$r^{2} e^{\sigma} F^{01} = 0$ is satisfied (see \cite{Bek71}).
In AR-VTG, without charges and currents, we can write $r^{2} e^{\sigma} F^{01} \equiv D$, where $D$ is a 
nonvanishing constant and, then, the field equations (\ref{fieles_vt}) 
lead to:
\begin{equation}
8\pi \Big[ \Big(\varepsilon - \frac {\gamma}{2} \Big) \frac{D^{2}}{r^{4}} - \rho u^{0} u_{0} \Big] =
\frac {1}{r^{2}} + e^{-\lambda} \Big( \frac {\lambda^{\prime}}{r} - \frac {1}{r^{2}} \Big) \ ,
\label{e_00}
\end{equation}    

\begin{equation}
8\pi \Big[ \Big(\varepsilon - \frac {\gamma}{2} \Big) \frac{D^{2}}{r^{4}} - \rho u^{1} u_{1} \Big] =
\frac {1}{r^{2}} - e^{-\lambda} \Big( \frac {\nu^{\prime}}{r} + \frac {1}{r^{2}} \Big) \ ,
\label{e_11}
\end{equation}    

\begin{equation}
8\pi \rho u^{1} u_{0} =
e^{-\lambda} \frac {\dot{\lambda}}{r} \ .
\label{e_01}
\end{equation}    
With the essential aim of matching metrics at the core surface, let us write the interior 
metric as follows:
\begin{equation}
e^{-\lambda} = 1 - \frac {2m(r,t)}{r} + 8\pi \Big(\varepsilon - \frac {\gamma}{2} \Big) \frac {D^{2}}{r^{2}} \ .
\label{math_m}
\end{equation}   
A similar procedure may be found in \cite{Bek71}. Since $D$ is a constant, 
this last equation plus Eq.~(\ref{e_00}) give:
\begin{equation}
\frac {\partial m}{\partial r} = -4 \pi \rho r^{2} u_{0} u^{0} \ ,
\label{der_m_r}
\end{equation}    
and, from Eqs.~(\ref{e_01}) and~(\ref{math_m}) one easily obtains
\begin{equation}
\frac {\partial m}{\partial t} = 4 \pi \rho r^{2} u_{0} u^{1} \ .
\label{der_m_t}
\end{equation}    
Finally, from the 4-velocity definition plus Eqs.~(\ref{der_m_r}) and~(\ref{der_m_t}) 
it follows that
\begin{equation}
\frac {d m}{d \tau} = \frac {\partial m}{\partial r} u^{1} +\frac {\partial m}{\partial t} u^{0} = 0 \ .
\label{der_m_tp}
\end{equation}    
According to this equation, the proper-time derivative of $m$ with respect to an 
observer comoving with the fluid vanishes, which means that the mass inside a 
sphere comoving with the collapsing matter is conserved. This is only true in the absence
of pressure. The same conservation holds in Einstein-Maxwell theory \citep{Bek71}. 
It is not possible for $\nabla \cdot A \neq 0$.
In particular, the total mass $M$ will be conserved since it is the mass inside the 
comoving boundary. This fact is necessary to match the interior metric (\ref{math_m})
and the exterior one; in fact, on the boundary one has $m(r,t) = M$ and 
by choosing 
\begin{equation}
8\pi \Big(\varepsilon - \frac {\gamma}{2} \Big) D^{2} = \alpha^{2} M^{2} \ ,
\label{ctants}
\end{equation}
the exterior metric has the same form as that defined by Eqs.~(\ref{LEK}) and~(\ref{rnm}). 

\subsection{The Basic Equations in comoving coordinates}
\label{sec:7b}

Hereafter, we will try to use comoving coordinates $(T,a,\theta,\phi)$ to 
get an analytical shell model inside the collapsing object; this will lead us to
a revision of the shell crossings  issue 
(an extensive analysis can be found in \cite{Hel85} within the framework of GR),
that is the collision of two adjacent dust shells. As stated in \cite{Ori90},
this fact is directly related to the absence of any coupling between the motion of
the different dust shells.
By using 
these coordinates, the line element inside the collapsing object has
the following form: 
\begin{eqnarray}
ds^{2}&=&\ - e^{\nu} dT^2 + e^{\lambda} da^2 \nonumber \\
& &
+ R^2 ( \ d\theta^2 + sin^2\theta \ d\phi^2) \ , 
\label{LEC}
\end{eqnarray}   
where $\nu$, $\lambda$, and $R > 0$ are functions of $a$ and $T$ (see \cite{Bek71,Lan75}).

For a collapsing spherically symmetric pressureless fluid, 
the conservation law $\nabla_{\mu} T^{\mu \nu}_{GR} = 0 $,
which is valid in GR as well as in AR-VTG (see Sect. \ref{sec:2}), leads to 
\begin{equation}
\dot{\lambda} + \frac {4\dot{R}} {R} = -2 \frac {\dot{\rho}} {\rho} \ ; \,\,\,\, \nu^\prime =0 \ .
\label{conlaw}
\end{equation} 
Hereafter, the prime (dot) denotes a derivative with repect to coordinate $a$ ($T$).
Since function $\nu $ does not depend on $a$ ($\nu^\prime =0$), the time $T$ may be 
redefined to have $\nu =0$, and the resulting coordinates are synchronous and comoving. 
In these coordinates the line element reads as follows \citep{Lan75,Ste04}:
\begin{eqnarray}
ds^{2}&=&\ - dT^2 + e^{\lambda} da^2 \nonumber \\
& &
+ R^2 ( \ d\theta^2 + sin^2\theta \ d\phi^2) \ . 
\label{LECS}
\end{eqnarray}   
Moreover, due to the remaining coordinate degree of freedom left,
comoving coordinate $a$ may be chosen in such a way that, inside the
collapsing core, it takes values inside the interval $0 \leq a \leq 1$, with $a=0$  
at the center and $a=1$ for the core boundary. 

By using the line element (\ref{LECS}), Eqs.~(\ref{1.3_vtm}) give:
\begin{equation}
(R^{2}e^{\lambda /2}F^{01})^{\prime} = 0
\label{1.3_vtm_c}
\end{equation} 
and 
\begin{equation}
(R^{2}e^{\lambda /2}F^{01})\dot{\,} = 0 \ .
\label{1.3_vtm_d}
\end{equation}  
These two equations express that quantity $R^{2}e^{\lambda /2}F^{01}$ is a constant.
Moreover,
in Schwarzschild coordinates, it has been proved that 
$r^{2} e^{\sigma} F^{01}$ quantity is constant too (see subsection \ref{sec:7b}). 
Both expressions give $\sqrt{-g}F^{01}$ in the corresponding coordinates.
Both constants are identical since $\sqrt{-g}F^{01}$ behaves as an scalar 
under coordinate transformations of the form $a=a(r,t)$ and $T=T(r,t)$
with fixed coordinates $\theta $ and $\phi $, which has the form of the 
transformations between 
comoving synchronous and Schwarzschild coordinates. Hence, 
in comoving synchronous coordinates, 
we can write 
\begin{equation}
R^{2}e^{\lambda /2}F^{01} = D \ , 
\label{com_syn_F}
\end{equation}  
and Eq.~(\ref{ctants}) holds. 
On account of these equations, plus the above expressions for
$T^{\mu \nu}_{GR}$ and $T^{\mu \nu}_{VT}$,
Eqs.~(\ref{fieles_vt}) may be written as follows 
\begin{eqnarray}
& & - 8\pi \Big[ \Big(\varepsilon - \frac {\gamma}{2} \Big) \frac {D^{2}}{R^{4}} + \rho \Big] =  \nonumber \\
& &
\frac {1}{R^{2}} \Big[ e^{-\lambda}(2RR^{\prime \prime} +
R^{\prime 2} - \lambda^{\prime}RR^{\prime})  \nonumber \\
& &                      
-(\dot{\lambda}R\dot{R}+\dot{R}^{2}) -1 \Big] \ ,
\label{eend_00}
\end{eqnarray}  
\begin{eqnarray}
& & - 8 \pi \Big[\Big(\varepsilon - \frac {\gamma}{2} \Big) \frac {D^{2}}{R^{4}} \Big] =  \nonumber \\
& &
\frac {1}{R^{2}} \Big[ e^{-\lambda}
R^{\prime 2} -(2R \ddot{R} + \dot{R}^{2}) -1 \Big] \ ,
\label{eend_11}
\end{eqnarray}  
\begin{equation}
- \frac{1}{R} e^{-\lambda} (2\dot{R}^{\prime} - \dot{\lambda}R^{\prime}) = 0 \ .
\label{eend_01}
\end{equation}  

As it is shown in \cite{Ste04}, the solution of Eq.~(\ref{eend_01}) 
is given by:
\begin{equation}
e^{\lambda} = \frac {R^{\prime 2}}{1 - \beta f^{2} (a)} \ ,
\label{ste_1}
\end{equation}   
where $\beta = 0, \pm 1$, and $f(a)$ is an 
arbitrary function of the radial coordinate $a$ subject the sole condition $1-\beta f^{2} (a) > 0$.
To make the reader easier, let us mention that the function $f$ we use and the used one by Landau \cite{Lan75}
and many other authors --say $f_L$--, are related by the expression $f_L = -\beta f^2$.

Substituting expression (\ref{ste_1}) in Eq.~(\ref{eend_11}), we find
\begin{equation}
2R\ddot{R} + \dot{R}^{2} = -\beta f^{2} (a) + \frac {\alpha^{2} M^{2}}{R^{2}} \ .
\label{ste_2}
\end{equation}    
This equation reduces to the corresponding equation of GR \citep{Ste04} for $\alpha^{2} =0$.
An integration gives 
\begin{equation}
\dot{R}^{2} = -\beta f^{2} (a) + \frac {F(a)}{R} -  \frac {\alpha^{2} M^{2}}{R^{2}} \ ,
\label{ste_2}
\end{equation}     
where $F(a)$ is a new arbitrary function. 

By combining Eqs.~(\ref{eend_00}), (\ref{ste_1}), and (\ref{ste_2}) one easily gets
the following formula for the fluid energy density of the $a$-shell at time $T$:
\begin{equation}
8\pi \rho(a,T) = \frac {F^{\prime}(a)} {R^{2}(a,T) R^{\prime}(a,T)} \ .
\label{denaT}
\end{equation}

Equation~(\ref{ste_2}) may be rewritten in the form
\begin{equation}
dT =  \frac {RdR}{\sqrt{-\beta R^{2} f^{2} (a) +RF(a) - \alpha^{2}M^{2}}} 
\label{ste_3}
\end{equation}   
and, evidently, the inequality 
\begin{equation}
P_{a}(R) \equiv \beta R^{2} f^{2} (a)-RF(a) + \alpha^{2}M^{2} < 0
\label{ineq}
\end{equation}     
must be satisfied whatever the radial coordinate $a$ value may be.

Assuming the weak energy condition ($T_{\mu \nu} V^{\mu} V^{\nu} \ge 0$) in GR,  it follows that $F' \ge 0$  \citep{Jos07}.
It is worth noting that the same conclusion can be drawn in AR-VTG from the expression $T^{\mu \nu} = T^{\mu \nu}_{GR} + T^{\mu \nu}_{VT}$.
In the same manner $F(a)$ can be interpreted as twice of the weighted mass (by the factor $\sqrt {1 - \beta f^{2}(a)}$ )
inside a volume $V$ of coordinate radius $a$, so $F$ must be positive everywhere ; $\beta f^{2} < 1$ must hold for a
Lorentzian manifold, as mentioned previously, in relation with  Eq.~(\ref{ste_1}).

\subsubsection{Elliptic Regions}
\label{sec:7b1}

Let us first consider $\beta = 1$. In such a case, the above inequality~(\ref{ineq}) requires $F(a) > 0$
for any given shell labeled by the comoving coordinate $a$. If functions $F(a)$ and $f(a)$ 
are chosen in such a way that 
\begin{equation}
\alpha^2 M^2 < F^2/4f^2 \ , 
\label{ineq1}
\end{equation}   
the condition $0 < F^{2}(a) - 4 f^{2}(a) \alpha^{2} M^{2} \equiv \Delta^{2} (a)$ is
satisfied for any $a$, and the
equation $P_{a}(R) = 0 $ has two real solutions:
\begin{equation}
R_{min} = [F(a) - \Delta(a)]/2f^{2}(a) > 0
\label{eqBeta1_Rmin}
\end{equation}
 and 
\begin{equation}
R_{max} = [F(a) + \Delta(a)]/2f^{2}(a) > 0.
\label{eqBeta1_Rmax} 
\end{equation}
It may be trivially verified that the inequality~(\ref{ineq}) is satisfied by $R$ values ranging inside the interval 
[$R_{min}$, $R_{max}$]. 
If a shell labeled as $a$ arrives to $R=R_{min}(a) $ 
at time $T=T_{0}(a)$, an integration of Eq.~(\ref{ste_3}) 
leads to
\begin{eqnarray}
T - {T_0}\left( a \right)&=&\  - \frac{1}{f^2}\sqrt { - {R^2}{f^2} + R\,F - {\alpha^{2}M^{2}}} \nonumber \\
& &
+ \frac{F}{f^3}{\sin ^{ - 1}}\sqrt {\frac{f^2 R}{\Delta } + \frac{\Delta  - F}{2\Delta}} \ . 
\label{eqBeta1_01}
\end{eqnarray}
Denoting $\chi \equiv \frac {2 \alpha M}{F} $, Eq.~(\ref{eqBeta1_01}) is written in the following parametric form:
\begin{equation}
\begin{aligned}
R = \frac{F}{{2{f^2}}}{[1- \sqrt {(1-{\chi}^2)}\cos \eta]}, \\
T - T_{0}(a) = \frac{F}{{2{f^3}}}{[\eta - \sqrt {(1-{\chi}^2)} \sin \eta]},
\label{eqBeta1_02}
\end{aligned}
\end{equation}
where $\eta$ is the parameter.

As it occurs with a test particle in the AR-VTG stationary
spherically symmetric background, inner shells could move between
two $R$ values, which are turning points satisfying the conditions
$\dot{R}(R_{max})=\dot{R}(R_{min})=0$ as it follows from Eq.~(\ref{ste_2}).
In order to avoid shell crossings in this motion, the arbitrary functions involved in the 
dynamic equations could  be chosen to ensure the condition 
$R^{\prime}(a,T)>0$, which guarantees (strictly enforced) no shell crossings \citep{Hel85,New86}.
This concern will be reviewed in the subsection \ref{sec:7c}.
\subsubsection{Parabolic Regions}
\label{sec:7b2}

For $\beta = 0$ and $F(a) >0$, inequality~(\ref{ineq}) reduces to $ R > \alpha^{2}M^{2} / F(a) $,
which means that any admissible $R$ value must be greater than

\begin{equation}
R_{min} =\alpha^{2}M^{2} / F(a).
\label{eqBeta0_Rmin}
\end{equation}

According to Eq.~(\ref{ste_2}), there is a turning point at $R=R_{min}$, where $\dot{R}$ vanishes.
Condition $F(a) < 0$ is not admissible, and now the resulting expression after integration of 
Eq.~(\ref{ste_3}) is:
\begin{equation}
T - T_{0}(a)  = 
\frac {2 \sqrt {FR-\alpha^{2} M^{2}}}{3F^{2}}
\Big( FR + 2 \alpha^{2} M^{2} \Big) \ . 
\label{eqBeta0_01}
\end{equation}

\subsubsection{Hyperbolic Regions}
\label{sec:7b3}

Finally, for $\beta = -1$, equation $P_{a}(R) = 0 $ has 
only a positive solution given by
\begin{equation}
R_{min} = [-F(a) + \tilde \Delta(a)]/2f^{2}(a) > 0,
\label{eqBeta-1_Rmin}
\end{equation}
 where 
$\tilde \Delta (a)^{2} \equiv F^{2}(a) + 4 f^{2}(a) \alpha^{2} M^{2} > 0$ for any $a$.
According to~(\ref{ineq}) any admissible $R$ value must be greater than 
$R_{min} $. By using Eq.~(\ref{ste_2}) one easily concludes that
$R=R_{min}$ is a unique turning point where $\dot{R}$ vanishes.
Now, the evolution equation for $R(a,T)$ takes the form
\begin{eqnarray}
T - {T_0}\left( a \right)&=&\ \frac{1}{f^2}\sqrt { {R^2}{f^2} + R\,F - {\alpha^{2}M^{2}}} \nonumber \\
& &
- \frac{F}{f^3}{\sinh ^{ - 1}}\sqrt {\frac{f^2 R}{\tilde \Delta } + \frac{F - \tilde \Delta}{2\tilde \Delta}} \ , 
\label{eqBeta-1_01}
\end{eqnarray}
or parametrically, can be written as:
\begin{equation}
\begin{aligned}
R = \frac{F}{{2{f^2}}}{\Big[\frac{\tilde \Delta }{F}\cosh \eta - 1\Big]}, \\
T - T_{0}(a) = \frac{F}{{2{f^3}}}{\Big[\frac{\tilde \Delta }{F}\sinh \eta - \eta \Big]}.
\label{eqBeta-1_02}
\end{aligned}
\end{equation}

Notice that the Eqs.~(\ref{eqBeta1_01})-(\ref{eqBeta-1_02}) describe an expanding or collapsing phase,
in accordance with $T \ge T(a)$ or $T \le T(a)$, respectively. We can parametrize this fact
replacing $T - T_{0}(a)$ by $\epsilon [T - T_{0}(a)]$ at Eqs.~(\ref{eqBeta1_01})-(\ref{eqBeta-1_02}),
where the new $\epsilon$ parameter values are:
$\epsilon = + 1$ for expansion and $\epsilon = - 1$ for the collapse.

\subsection{Shell Crossings}
\label{sec:7c}

For $\beta = 0$ and $\beta = -1$ just one turning point exists,
and oscillations of shells are not possible. There is only a minimum 
$R_{min} $ and, consequently, falling shells bounce at $R=R_{min} $ and then    
expand forever. 

The above discussion of cases $\beta = 1$ (two turning points), $\beta = 0$ (one) 
and $\beta = -1$ (one) is consistent with the behavior of a test particle in the fixed AR-VTG stationary
spherically symmetric background, which has two turning points for $E < 1/2 $ and only one for
$E \geq 1/2 $. Let us discuss separately the existence of shell crossings for $\beta =1$,
$\beta =0$, and $\beta =-1$.

In the first case, using the following expression for $R'$ 
--which is obtained from Eqs.~(\ref{eqBeta1_02})-- , 

\begin{eqnarray}
R' &=& \left( {\frac{{F'}}{F} - \frac{{2f'}}{f}} \right)R \nonumber \\
& &
- \left[ {{T_0}^\prime  + \left( {\frac{{F'}}{F} - \frac{{3f'}}{f}} \right)\left( {T - {T_0}} \right)} \right]\dot R \nonumber \\
& &
+ \frac{2 \alpha^2 M^2}{\Delta^2}\left( {\frac{{F'}}{F} - \frac{{f'}}{f}} \right)\left( {F - \frac{{2 \alpha^2 M^2}}{R}} \right)
\label{eqBeta1_03}
\end{eqnarray}
it can be proved that, 
for any shell, quantities  $R'[R_{min}(a)]$  and $R'[R_{max}(a)]$
have opposite signs and, consequently, $R'[R(a,T)]$ vanishes 
for some $R$ between $R_{min}(a)$ and $R_{max}(a)$; hence, for $\beta=1 $
the use of comoving coordinates numbering shells is not a good choice to fully describe the internal shells collapse.
A similar situation it is analyzed in \cite{Ori91} and \cite{Kra06} for a spherically symmetric charged dust collapse,
and also in \cite{Kra12} where the charged dust solution of Ruban \cite{Rub72}, 
a generalisation of the Datt \cite{Dat38} solution for a non-charged dust is considered. It can be found in \cite{Gon01} that, 
in spite of the inclusion of a positive cosmological constant, this repulsion does not prevent the shell crossings.  
However, it is possible to describe a first bounce if the comoving time $T$ origin is not chosen at $R_{max}(a)$, say $R{_N}(a)$,
in which case, a set of functions $F(a)$, $f(a)$ and $T_0(a)$ so that $R'(a) > 0$, 
$F(a) > 0$ and $F'(a) > 0$ inside the interval [$R_{N}$, $R_{min}$] can be found, that is, our picture has
begun with a certain $\dot R \ne 0$; in such a case the following inequality have to be satisfied:

\begin{equation}
\frac{F'}{F} < \frac{f'}{f} \Big(1 - \frac {\Delta}{F}\Big) . 
\end{equation}

When the $\beta = 0$ case is considered, taking the derivative of Eq.~(\ref{eqBeta0_01}) it follows that
\begin{equation}
\begin{aligned}
R' = \frac{1}{3}\frac {F'}{F}R + \frac{4}{3}\frac{F'}{F^2} \alpha^2 M^2 \left( {1 - 2\frac {\alpha^2 M^2}{FR}} \right) - {T'}_0 \dot R,
\label{eqBeta0_02}
\end{aligned}
\end{equation}
and therefore
\begin{equation}
 \lim \limits_{T \to {T_{0}(a)}}R'(a,T) = - {F'\alpha^{2} M^{2}} / {F^2} = R'_{min}(a) . \nonumber
\end{equation}
From Eq.~(\ref{denaT}) it is immediately evident that
\begin{equation}
\lim \limits_{T \to {T_{0}(a)}}8 \pi \rho(a,T) = - {F^4} / \alpha^{6} M^{6} = 8 \pi \rho_{min}(a) < 0. \nonumber
\end{equation}
So, if the collapse was started at certain $R = R{_N}(a)$ with a certain energy density for the $a$-shell defined by
$8 \pi \rho{_N}(a) = F'(a) / R_{N}^{2}(a)  R'_{N}(a) > 0$, which means that the sign of $F'(a)$ and  $R'_{N}(a)$
is the same, then the sign of $R'(a,T)$ will change at $R_{min}$. Hence
we can conclude that the shell crossings is unavoidable (see section IIb at reference \cite{Hel85}).

And finally, let us see that in the case $\beta = -1$, the arbitrary functions involved in the model 
may be chosen to prevent shell crossings. In order to achieve this aim, first of all $R'$ is calculated using the set of Eqs.~(\ref{eqBeta-1_01})
which leads to

\begin{eqnarray}
R' &=& \left( {\frac{{F'}}{F} - \frac{{2f'}}{f}} \right)R \nonumber \\
& &
- \left[ {{T_0}^\prime  + \left( {\frac{{F'}}{F} - \frac{{3f'}}{f}} \right)\left( {T - {T_0}} \right)} \right]\dot R \nonumber \\
& &
+ \frac{2 \alpha^2 M^2}{\tilde \Delta^2}\left( {\frac{{F'}}{F} - \frac{{f'}}{f}} \right)\left( {F - \frac{{2 \alpha^2 M^2}}{R}} \right).
\label{eqBeta-1_03}
\end{eqnarray}
Then taking into account the condition $R'[R_{min}(a)] > 0$, that is accomplished iff the inequality
\begin{equation}
\frac{F'}{F} < \frac{f'}{f} \Big(1 - \frac {\tilde \Delta}{F}\Big) 
\label{eqBeta-1_03}
\end{equation}
is satisfied for any $a$-shell, we may select an appropriate function $T_{0}(a)$  
in order to guarantee $R'(a,T) > 0$ in the interval [$R_{N}$, $R_{min}$[.  In the next section, other
considerations will be taken into account which will limit this result.

For a general solution, numerical methods and codes similar to those used to study the collapse in GR 
\citep{Fon03,Ghe05,Ghe07,Oco10,Ger16} have to be adapted to AR-VTG. Nevertheless, this is a
hard task which is out of this paper's scope.

\subsection{Apparent Horizons and Trapped Surfaces}
\label{sec:7d}

Now we analyze the apparent horizons for the AR-VTG, that reveal the boundary of the trapped surfaces
specifying the region from which no light is allowed to escape.
In order to obtain, the condition $g^{\mu \nu }{\partial _\mu }R{\partial _\nu }R = 0$ and the Eqs.~(\ref{ste_1}) and (\ref{ste_2}) are used.
The table below (Table~\ref{tab:01}) shows the three different cases that can happen.

As it can be appreciated, the existence and number of apparent horizons, do not depend on the local type
of time evolution defined by $\beta$ value, furthermore do not depend on the particular form of the $f$ function. The $R = F(a)$ horizon
of GR is recovered when setting $\alpha = 0$, as expected.

\begin{table}[ph]

\caption{\label{tab:01} AR-VTG apparent horizons in comoving coordinates. For simplicity we define ${\cal M}(a) \equiv F(a)/2$.}
\begin{ruledtabular}
\begin{tabular}{ccc}
Condition & Number of  horizons &  $R$ Value \\ 
\hline \\

${\cal M} < \alpha M$ & No horizon & Not applicable    \\ 

${\cal M} = \alpha M$ & One horizon & $R_{H} \equiv {\cal M} = \alpha M$    \\ 

${\cal M} > \alpha M$ & Two horizons & $R_{H_ \pm} \equiv {\cal M} \pm \sqrt {{{\cal M}^2} - \alpha^{2} M^{2}}$    \\ 

\end{tabular}
\end{ruledtabular}
\end{table}

In the following, we present the different situations, depending on $\beta$ value, for the last entry at Table~\ref{tab:01},
that is, when there are two apparent horizons. Firstly, for $\beta = 0$, taking into account Eqs.~(\ref{eqBeta0_Rmin})-(\ref{eqBeta0_01})
and the $R_{H_ \pm}$ definition, is easy to derive the inequalities $R_{min} < R_{H_{-}} < R_{H_{+}}$ and $T_{0} > T_{H_{-}} > T_{H_{+}}$, where
$T_{H_ \pm}$ represents the time when the $a$-shell crosses the horizon $R_{H_ \pm}$; the same results are obtained in the case of $\beta = -1$ when
considering Eqs.~(\ref{eqBeta-1_Rmin})-(\ref{eqBeta-1_01}). And secondly, for $\beta = 1$, in which case two turning points exist, 
using Eqs.~(\ref{eqBeta1_Rmin})-(\ref{eqBeta1_01}) together with the fact that condition $f^{2} < 1$ must be satisfied, the inequalities $R_{min} < R_{H_{-}} < R_{H_{+}} < R_{max}$ 
and $T_{0} > T_{H_{-}} > T_{H_{+}} > T_{max}$ are obtained, where $T_{max} = T_0 - \pi F/2f^3$ represents time $T$ value for a $a$-shell at $R_{max}$. In this case,
the condition for the last row at Table~\ref{tab:01} with the already mentioned inequality $f^2 < 1$, provides the restriction (\ref{ineq1}) to be fulfilled. 
So, in any case, any signal emitted  from $R_{min}$ (that is at $T=T_0$) is future trapped at the region above ${T=T_{H_{-}}}$. \\

\section{Discussion and conclusions}   
\label{sec:8}

Our starting point has been the vacuum static spherically symmetric metric of AR-VTG given by Eqs.~(\ref{LEK}) and (\ref{rnm}).
This metric and some values of $r_{BH}$ corresponding to $\alpha^2 < 1$ were found in \cite{Dal15}, where the singularity was
not considered at all. A very different behavior to GR is found in the introduction of Sect. \ref{sec:7} where a test particle on the 
AR-VTG fixed background is presented; it is revealed that oscillatory trajectories for the aforementioned particles can be obtained. One of the first 
published descriptions of the motion of a point on the surface of a charged collapsing ball 
of mass $M$ and charge $Q$ can be found at \cite{Nov66}. This description is fully equivalent to our case, however in AR-VTG
gravity no charge is necessary to obtain a similar motion. The $\alpha^2 < 1$ condition in AR-VTG is analogous to charge to the mass ratio one
${(Q/M)^2 < 1}$ in the Reissner-Nordstr\"{o}m spacetime, where the surface of the collapsing sphere crosses the inner horizon
${r_{inh} = M - \sqrt{M^2 - Q^2}}$ (in AR-VTG ${r_{H_{-}} = M(1 - \sqrt{1 - \alpha^2})}$). An equivalent spacetime diagram 
can be found at \cite{Kra06B}.

The dynamic properties of AR-VTG are presented in Sect. \ref{sec:7}, divided in different subsections, were the dust core is considered. 
The study of the internal core in comoving coordinates leads to, as in GR, three possible families of solutions 
that classify the spacetime as bound, marginally bound or unbound \cite{Jos07}. 
This fact has been characterized with $\beta = 1$, $\beta = 0$ and $\beta = -1$ respectively.
In all three cases a minimum value for ${R}$ has been found, say ${R_{min}(a,T_0(a))}$,  where ${\dot R(a,T_0(a)) = 0}$ and which is reached after crossing the apparent horizon at ${T_{H_{-}}}$, then the collapse is halted and reversed.
The same fact was first found for a non charged dust in an external electromagnetic field by Shikin \cite{Shi72}.
So, under the premise that the real Universe and astrophysical objects (in general) have no net electric charge, 
AR-VTG provides an interesting alternative for preventing gravitational singularities as a pure classical gravitation effect.

The shell crossings issue has been analyzed, and several similarities can be observed when a charged dust is considered
(see Sections VI and VII in \cite{Kra06}). However there are also differences, for instance, the absence of real charges eliminates fundamental equations relative to these, and changes some
of the regularity conditions at the center of symmetry as the fact that charge has to be zero there.
In AR-VTG, while using comoving coordinates, we have found the unavoidable shell crossings for $\beta = 0$, for any collapsing $a$-shell before reaching $R_{min}(a)$. 
So, the use of comoving coordinates does not allow the possibility to follow further evolution. For $\beta = 1$ it is possible to find a set of appropiate 
free functions that should allow us to avoid shell crossings collapse during a first bounce at $R_{min}$, although it will be unavoidable during the expansion
before reaching $R_{max}$. Finally, when $\beta = -1$ it has been found, apparently, a configuration (defined by the inequality (\ref{eqBeta-1_03})) avoiding the issue, but there are extra constraints to be considered due to the regularity at the symmetry center. From inequality (\ref{eqBeta-1_03}) it is immeditly derived that the quotient $f'/f$ has to be negative. Therefore
one of the conditions related as necesary at \cite{Vic73} for a charged dust,  which is $f(a_c) = 0$ (where $a = a_c$ is the symmetry center), cannot be honored.
Nevertheless, an initial cloud with a Minkowskian central bubble should not require the $f(a_c) = 0$ condition \cite{Ori91}, and then inequality (\ref{eqBeta-1_03}) may be achieved.
Following the same procedure that Krasi\'nski and Bolejko present in \cite{Kra06}, it can be demonstrated that also in AR-VTG gravity shell crossings is a coordinate singularity, so inherent to the coordinates selection. According to \cite{New86} the degeneracy created in the metric (\ref{LECS}) when $R' = 0$ can be saved through a $C^0$ extension of it. Once the fluid has been modeled into shells, this process represents the collisions of adjacent shells and it is related to the fact that there is no coupling between the movement of different shells.  

The fundamental symmetry of AR-VTG is ${A^\mu \to A^\mu + \nabla ^\mu \Phi}$ with ${\nabla_ \mu \nabla ^\mu \Phi = 0}$, which is different from the standard U(1) gauge symmetry \cite{Dal14}; the quantities ${\nabla  \cdot A}$ and ${F_{\mu \nu}}$ are invariant under these tranformations.
In cosmology, a quantity proportional to ${(\nabla \cdot A)^2}$ is a candidate to play the role of dark energy with an equation of state $w = p/\rho= -1$, and AR-VTG has been demonstrated to be in agreement with the background and perturbed spacetimes \citep{Dal09,Dal14,Dal17}, while ${F_{\mu \nu}}$ vanishes.
In this paper it has been shown that a quantity related with ${F_{01}}$ plays the role of the repulsive component of the gravitation
and may prevent the gravitational collapse, while ${\nabla \cdot A}$ is null.

The influence of the repulsive component of AR-VTG on neutron star structure 
is also an open problem deserving attention. The presented dust collapse model provides an interesting alternative of a non singular collapse as 
may be the weakly charged dust. An interesting open issue is the study of the $A_\mu$ field perturbations that may save the unstable character found 
in a charged black hole where the innermost sections are unstable to electromagentic perturbations \cite{Ori91}. 

\begin{acknowledgments}
This work has been supported by the Spanish
``Ministerio de Econom\'{\i}a y Competitividad'' and the ``Fondo Europeo de Desarrollo 
Regional'' MINECO-FEDER Project No. FIS2015-64552-P. We thank J. A. Morales-Lladosa for
useful discussion and support.
 
\end{acknowledgments}

\end{document}